\documentclass[11pt,epsf,amssymb,qsymbols]{article}
\usepackage{tabularx}
\usepackage{array}
\usepackage{graphics}
\usepackage{graphicx}
\usepackage{epsfig}
\usepackage{amsmath}
\usepackage{amssymb}
\usepackage{psfrag}
\usepackage{color}
\usepackage{longtable}
\makeatletter
\usepackage{verbatim}

\normalsize
\newcommand{\bea}{\begin{eqnarray}}
\newcommand{\eea}{\end{eqnarray}}
\newcommand{\beq}{\begin{equation}}
\newcommand{\eeq}{\end{equation}}

\newcommand{\pdir}{p\kern -5.2pt\raise 0.2ex\hbox {/}}
\newcommand{\vdir}{v\kern -5.75pt\raise 0.15ex\hbox {/}}
\newcommand{\kdir}{k\kern -5.75pt\raise 0.15ex\hbox {/}}
\newcommand{\epsdir}{\epsilon\kern -5.0pt\raise 0.15ex\hbox {/}}
\newcommand{\bvdir}{\bar{v}\kern -5.75pt\raise 0.15ex\hbox {/}}
\newcommand{\Ddir}{D\kern -7.75pt\raise 0.20ex\hbox {/}}
\newcommand{\ldir}{l\kern -5.0pt\raise 0.2ex\hbox{/}}
\newcommand{\varepsdir}{\varepsilon\kern -5.5pt\raise 0.15ex\hbox{/}}

\makeatother
\def\slashed#1{\gamma\cdot #1}

\definecolor{Red}{rgb}{1.,0.,0.}
\newcommand{\Red}[1]{{\color{Red}{#1}}}

\begin{document}

\thispagestyle{empty}
\begin{center}
\vskip 2.7cm\par
{\par\centering \Large \bf Chiral loop corrections to weak decays of $B$ mesons to positive and negative
parity charmed mesons}\\
\vskip 1.75cm\par
\scalebox{.89}{\par\centering \large
\sc Jan O. Eeg$^a$, Svjetlana~Fajfer$^{b,c}$ and Jernej~Kamenik$^{b}$}
{\par\centering \vskip 0.5 cm\par}
{\sl
$^a$ University of Oslo, Physics Department, \\
P. O. Box 1048 Blindern, N-0316 Oslo 3, Norway.\\
\vspace{.25cm}
$^b$ J.~Stefan Institute, Jamova 39, P.O. Box 3000,\\
1001 Ljubljana, Slovenia.\\
\vspace{.25cm}
$^c$
Department of Physics, University of Ljubljana,\\
 Jadranska 19, 1000
Ljubljana,
Slovenia.  }\\
{\vskip 0.25cm \par}
\end{center}

\vskip 1.25cm
\begin{abstract}
We determine chiral loop corrections to the $B$ meson decay amplitudes to positive and negative parity charmed mesons within a framework which combines heavy quark and chiral symmetries.  Then we investigate the impact of the lowest-lying positive parity heavy mesons on the determination of the Isgur-Wise functions. The corrections due to these states are competitive with the contributions arising from $K$ and $\eta$ meson loops. Since lattice studies rely on the chiral behavior of the amplitudes we discuss the chiral limit of our results. We find that the determination of the slope at zero recoil of the Isgur-Wise function $\xi$ for the $B$ transition to negative parity charm mesons is moderately affected by the inclusion of new states, while the slope of $\tau_{1/2}$ is affected significantly more.
\end{abstract}
\vskip 0.2cm
\setcounter{page}{0}
\setcounter{footnote}{0}
\setcounter{equation}{0}
\noindent

\renewcommand{\thefootnote}{\arabic{footnote}}

\newpage
\setcounter{footnote}{0}

\section{Introduction}
The present experimental knowledge on the values of the CKM
parameters is firmly established. The goal is to achieve highest
possible accuracy in their experimental extraction. In order to
reach it in exclusive decay modes one has to gain maximal control
over the relevant form factors. In the case of precise determination
of the $V_{cb}$ CKM matrix elements the studies of $B$ meson decays
into charm resonances have been playing a prominent role. In
experiments aimed to determine $V_{cb}$, actually the product
$|V_{cb} {\cal F } (1) |$ is extracted, where  ${\cal F } (1)$ is
the $B \to D$ or $B \to D^*$ hadronic form factor at zero recoil. A
lack of precise information about the shapes of various form factors
is thus still the main source of uncertainties. In theoretical
studies, heavy quark symmetry has been particularly appealing due to
the reduction of six form factors in the case of $B \to D(D*) l
\nu_l$ transitions to only one~\cite{Isgur:1989ed,Isgur:1990jf}. In
addition, at zero recoil, when the final state meson is at rest in
the $B$ rest frame, the normalization of the form factors is fixed
by symmetry. However, the results obtained within heavy meson
effective theories obtain important corrections coming from
operators which are suppressed as $1/M_{B,D}$~\cite{Boyd:1995pq} as
well as of higher order in the chiral
expansion~\cite{Cho:1992cf,Falk:1993iu,Casalbuoni:1996pg,Jenkins:1992qv}.
The knowledge of both kinds of corrections has improved during the
last few years.  The $B \to D^* l \nu_l$ decay amplitude is
corrected by  $1/M_{B,D}$ only at the second order in this expansion
making it more appropriate for the experimental
studies~\cite{Casalbuoni:1996pg,Luke:1990eg}. In addition to heavy
meson effective theory, other approaches have been used in the study
of the $B \to D (D^*)$ form factors, such as quark
models~\cite{DeVito:2006sq} and QCD sum rules~\cite{Dai:1998ca},
while the most reliable results should be expected from lattice
QCD~\cite{Kronfeld:2003sd}. In the treatment of hadronic properties
using lattice QCD the main problems arise due to the small masses of
the light quarks. Namely, lattice studies have to consider light
quarks with larger masses and then extrapolate results to their
physical values. In these studies the chiral behavior of the
amplitudes is particularly important. Heavy meson chiral
perturbation theory (HM$\chi$PT) is very useful in giving us some
control over the uncertainties appearing when the chiral limit is
approached~\cite{Becirevic:2006me,Fajfer:2006hi}. Most recently in
ref.~\cite{Laiho:2005ue}, the authors have discussed $B \to D l
\nu_l$ and $B \to D l \nu_l$  form factors in staggered chiral
perturbation theory by including next-to-leading order corrections
in staggered chiral perturbation theory.

\par

The practitioners of the heavy meson effective theories faced new
tasks when the charm mesons of the positive parity were discovered.
In addition to understanding their structure, mass differences and
decay properties the fact that observed resonances lie  only about
$350~\mathrm{MeV}$ above negative parity states stimulated many
studies~\cite{Colangelo:2004vu,Becirevic:2004uv,Lin:2006vc}. The
inclusion of these states into the heavy meson effective theory was
done fifteen years ago by including a number of unknown
parameters~\cite{Cho:1992cf,Falk:1993iu,Jenkins:1992qv} into the
HM$\chi$PT Lagrangian. Recently,
in~\cite{Becirevic:2006me,Fajfer:2006hi} the role of positive parity
states in the $B_{d,s}^0 - \bar B_{d,s}^0$ transitions and strong
decays was investigated and a few important statements were reached:
the contributions coming from positive parity states are competitive
in size with the kaon and $\eta$ meson loop corrections. However,
they  do not alter the pion chiral logarithms and consequently they
provide a guideline for the lattice extrapolation of these results.
A similar conclusion was already hinted on long time ago in the case
of $B\to D^{(*)}$ form factors~\cite{Falk:1993iu}, although a complete
analysis could not be performed at the time.

\par

In this paper we reinvestigate chiral loop corrections within
HM$\chi$PT to the semileptonic transitions of $B$ mesons into charm
mesons of negative as well as positive parity. Specifically we study
the effects of the small mass splitting between positive and
negative parity heavy meson states on the leading non-analytic
chiral behavior of the amplitudes. In Sec. 1 we give the main
details of our framework. Sec. 3 contains calculation of chiral
loops while in Sec. 4 we discuss chiral extrapolation. In Sec. 5 we
briefly summarize our results.

\section{Framework}

We use the formalism of heavy meson chiral
Lagrangians~\cite{Burdman:1992gh,Wise:1992hn}. The octet of light
pseudoscalar mesons can be encoded into $\Sigma = \xi^2 =
\mathrm{exp} (2i \pi^i \lambda^i /f)$ where the $\pi^i \lambda^i$
matrix contains the pseudo-Goldstone fields
\begin{equation}
\pi^i\lambda^i =
   \begin{pmatrix} \frac{1}{\sqrt 6}\eta + \frac{1}{\sqrt 2} \pi^0 & \pi^+ & K^+ \\
   \pi^- & \frac{1}{\sqrt 6}\eta - \frac{1}{\sqrt 2} \pi^0 & K^0 \\
   K^- & \overline K^0 & -\sqrt{\frac{2}{3}}\eta \end{pmatrix}
\end{equation}
and $f\approx 120~\mathrm{MeV}$ at one loop~\cite{Gasser:1984gg}.
The heavy-light mesons are customarily cataloged using the total
angular momentum of the light degrees of freedom in the heavy meson
$j_{\ell}^P$ which is a good quantum number in the heavy quark limit
due to heavy quark spin symmetry. The negative $(j_{\ell}^P)=1/2^-$
and positive $(j_{\ell}^P)=1/2^+$ parity doublets can then be
respectively represented by the fields $H(v)=1/2(1+\slashed v
)[P^{*}_{\mu}(v) \gamma^{\mu} - P(v) \gamma_5]$, where
$P^{*}_{\mu}(v)$ and $P(v)$ annihilate the vector and pseudoscalar
mesons of velocity $v$, and $S(v)=1/2(1+\slashed v )[P^{*}_{1\mu}(v)
\gamma^{\mu}\gamma_{5} - P_0(v)]$ for the axial-vector
($P^{*}_{1\mu}(v)$) and scalar ($P_0(v)$) mesons.
\par
The strong interactions Lagrangian relevant for our study of chiral
corrections to processes among heavy mesons of velocity $v$ is then
at leading order in chiral and heavy quark expansion
\begin{eqnarray}
    \mathcal L &=& \mathcal L_{\chi} + \mathcal L_{\frac{1}{2}^-} + \mathcal L_{\frac{1}{2}^+} + \mathcal L_{\mathrm{mix}}, \nonumber\\
    \mathcal L_{\chi} &=& \frac{f^2}{8} \partial_{\mu} \Sigma_{ab} \partial^{\mu} \Sigma^{\dagger}_{ba} + \lambda_0 \left[(m_q)_{ab} \Sigma_{ba} + (m_q)_{ab} \Sigma^{\dagger}_{ba}\right], \nonumber\\
    \mathcal L_{\frac{1}{2}^-} &=& - \mathrm{Tr}\left[ \overline H_a(v) (i v \cdot \mathcal{D}_{ab} - \delta_{ab} \Delta_H ) H_b(v)\right]  + g \mathrm{Tr} \left[ \overline H_b(v) H_a(v) \slashed{\mathcal A}_{ab} \gamma_{5} \right], \nonumber\\
    \mathcal L_{\frac{1}{2}^+} &=& \mathrm{Tr} \left[\overline S_a(v) ( i v \cdot \mathcal{D}_{ab} - \delta_{ab} \Delta_S) S_b(v) \right]  + \tilde g \mathrm{Tr} \left[ \overline S_b(v) S_a(v) \slashed{\mathcal A}_{ab} \gamma_{5}  \right], \nonumber\\
    \mathcal L_{\mathrm{mix}} &=& h \mathrm{Tr} \left[ \overline H_b(v) S_a(v) \slashed{\mathcal A}_{ab} \gamma_{5}  \right] + \mathrm{h.c.}.
    \label{L_1}
    \label{eq_L_0}
\end{eqnarray}
$\mathcal{D}^{\mu}_{ab} = \delta_{ab}\partial^{\mu}
-\mathcal{V}^{\mu}_{ab}$ is the covariant heavy meson derivative.
The light meson vector and axial currents are defined as $\mathcal
V_{\mu } = 1/2 (\xi^{\dagger} \partial_{\mu} \xi + \xi
\partial_{\mu} \xi^{\dagger})$ and $\mathcal A_{\mu } = i/2
(\xi^{\dagger} \partial_{\mu} \xi - \xi
\partial_{\mu}\xi^{\dagger})$ respectively. A trace is taken over
spin matrices and the repeated light quark flavor indices.
The $\lambda_0$ term induces masses of the pseudo-Goldstone mesons $m^2_{ab}= 4 \lambda_0 (m_a + m_b)/f^2$.
Accordingly $\mathcal L_{\chi}$ is of the order $\mathcal O (p^2)$
in the chiral power counting while the rest of this leading order
Lagrangian is of the order $\mathcal O(p^1)$.  Exceptions are the
$\Delta_H$ and $\Delta_S$ residual masses of the $H$ and $S$ fields
respectively. In a theory with only $H$ fields, one is free to set
$\Delta_H=0$ since all loop divergences are cancelled by $\mathcal
O(m_q)$ counterterms at zero order in $1/m_H$ expansion. However,
once $S$ fields are added to the theory, another dimensionful
quantity $\Delta_{SH} = \Delta_S - \Delta_H$ enters loop
calculations and does not vanish in the chiral and heavy quark
limit~\cite{Mehen:2005hc}.
We fix its value close to the phenomenological mass splitting
between the even and odd parity heavy meson multiplets $\Delta_{SH}
\approx 400~\mathrm{MeV}$ although smaller values have also been
proposed when taking into account next to leading order terms in
$1/m_H$ expansion~\cite{Mehen:2005hc}. For the couplings $g$, $h$ and $\tilde g$ we use the recently estimated values of~\cite{Fajfer:2006hi} $g \simeq 0.6$, $h \simeq -0.5$ and $\tilde g \simeq -0.1$.
\par
The weak part of the Lagrangian describing transitions among heavy
quarks can be matched upon weak heavy quark currents in
HQET~\cite{Falk:1993iu,Wise:1992hn}
\begin{eqnarray}
\overline c_{v} \Gamma b_{v'} &\to& C_{cb} \Huge\{ -\xi(w) \mathrm{Tr} \left[ \overline{H}_a(v) \Gamma H_a(v') \right] \nonumber\\
&&\hskip-1cm-\tilde \xi(w) \mathrm{Tr} \left[ \overline{S}_a(v) \Gamma S_a (v') \right] -\tau_{1/2}(w) \mathrm{Tr} \left[ \overline{H}_a(v) \Gamma S_a (v') \right] + \mathrm{h.c.} \Huge\}
\end{eqnarray}
at leading order in chiral and heavy quark expansion and where
$\Gamma=\gamma_{\mu}(1-\gamma_5)$ and $w= v \cdot v'$. Note that heavy quark symmetry dictates the values of $\xi(1)=\tilde \xi(1) =1$, which should not receive any chiral corrections. On the other hand $\tau_{1/2}(w)$ is not constrained and we use the recently determined value of~\cite{Becirevic:2004ta} $\tau_{1/2}(1)=0.38$.

\section{Calculation of Chiral Loop Corrections}

\begin{figure}[!t]
\begin{center}
\epsfysize2.3cm\epsffile{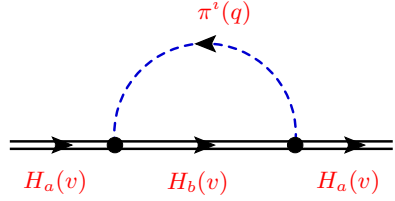}
\end{center}
\caption{\label{diagram_sunrise}{\footnotesize\sl Self-energy ("sunrise" topology) diagram. Double lines represent heavy mesons, dashed lines represent pseudo-Goldstone bosons, while filled dots represent effective strong vertices.}}
\end{figure}

\begin{figure}[!t]
\begin{center}
\epsfysize2.3cm\epsffile{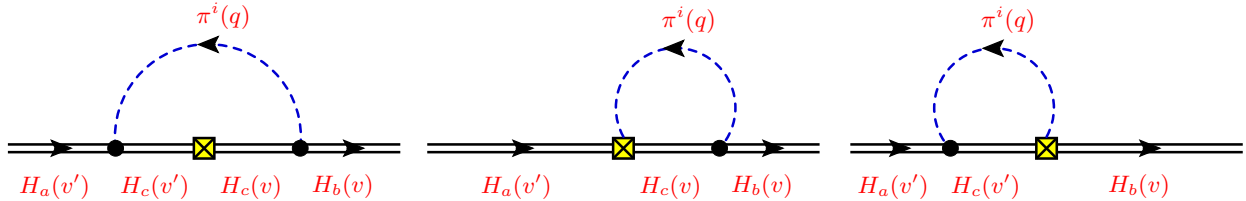}
\end{center}
\caption{\label{diagram_xi}{\footnotesize\sl Weak vertex correction diagrams. Crossed boxes represent effective weak vertices. Only diagrams of the outmost left topology contribute to the amplitude at the leading chiral log order.}}
\end{figure}

Here we present the most important details of our calculation of
leading chiral loop corrections to the Isgur-Wise functions $\xi(w)$, $\tilde \xi(w)$ and
$\tau_{1/2}(w)$. Following Refs.~\cite{Boyd:1994pa,Stewart:1998ke}, we
absorb the infinite and scale dependent pieces from one loop
amplitudes into the appropriate counterterms at order $\mathcal O
(m_q)$ (see e.g.~\cite{Fajfer:2006hi}). We first calculate the wave
function renormalization $Z_{2H}$ of the heavy $H(v)=P(v),~P^*(v)$
and $P_0(v)$, $P^*_1(v)$ fields. This has been done e.g. in
Ref.~\cite{Fajfer:2006hi} and we only quote the result at the
$\mathcal O(p^2)$ power counting order (in Appendix~A). We get
non-zero contributions to the heavy meson wavefunction
renormalization from the self energy ("sunrise" topology) diagrams
in Figure~\ref{diagram_sunrise} with leading order couplings in the
loop. \psfrag{pi}[bl]{\footnotesize $\Red{ {\pi^i(q)}}$}
\psfrag{Ha}[cc]{\footnotesize $\Red{H_a(v)}$}
\psfrag{Hb}[cc]{\footnotesize $\Red{H_b(v)}$}
\psfrag{Pstara}[bc]{$P^{*\mu}_a(v)$}
\psfrag{Sstarb}[bc]{$P^{*\nu}_{1 b}(v)$} \psfrag{Sa}[bc]{$P_{0
a}(v)$} \psfrag{Sstara}[bc]{$P^{*\mu}_{1 a}(v)$}
In the case of the $P(v)$ mesons both vector $P^*(v)$ and scalar
$P_0(v)$ mesons can contribute in the loop. The positive parity
$P_0(v)$ and $P^*_1(v)$ similarly obtain wavefunction
renormalization contributions from self energy diagrams
(Figure~\ref{diagram_sunrise}) with $P^*_1(v)$, $P(v)$ and $P_0(v)$,
$P^*_1(v)$, $P^*(v)$ mesons in the loops respectively. Then we
calculate loop corrections to the effective weak vertices. These
come from the one loop diagram topologies shown in
Figure~\ref{diagram_xi}. \psfrag{pi}[bl]{\footnotesize
$\Red{\pi^i(q)}$} \psfrag{Ha}[cc]{\footnotesize $\Red{H_a(v')}$}
\psfrag{Hb}[cc]{\footnotesize $\Red{H_b(v)}$}
\psfrag{Hc}[cc]{\footnotesize $~\Red{H_c(v')}$}
\psfrag{Hd}[cc]{\footnotesize $~~\Red{H_c(v)}$}
Namely, the initial and final heavy states may exchange a
pseudo-Goldstone, while pairs of positive and negative parity heavy
mesons may propagate in the loop (the left diagram in
Figure~\ref{diagram_xi}). Again not all heavy states contribute due to
parity conservation in effective strong interaction vertices. Thus, when initial and final states are pseudoscalars we get contributions from pairs of $P^*(v')P^*(v)$, $P_0(v')P^*(v)$, $P^*(v')P_0(v)$ and $P_0(v')P_0(v)$ propagating in the loop, while for pseudoscalar initial and scalar final state we get contributions from pairs of $P^*(v')P(v)$, $P^*(v')P_1^*(v)$, $P_0(v')P(v)$ and $P_0(v')P_1^*(v)$ in the loop (due to heavy quark symmetry, the same results are obtained for (axial)vector external states, although different intermediate states contribute). On the
other hand diagrams containing pseudo-Goldstone emission or
absorption from the weak vertex (central and right diagram in
Figure~\ref{diagram_xi}) do not contribute at all at the leading order
in the chiral power counting, since they may only appear via new weak operators containing derivatives or mass operators of the pseudo-Goldstone fields.  The complete expressions for the loop
corrected $\xi(\omega)$, $\tilde \xi (\omega)$ and
$\tau_{1/2}(\omega)$ we obtain are rather lengthy and can be found
in Appendix~A.

\section{Chiral Extrapolation}

We study the contributions of the additional resonances in the
chiral loops to the chiral extrapolations employed by lattice QCD
studies to run the light meson masses from the large values used in
the simulations to the chiral limit~\cite{Abada:2003un,
McNeile:2004rf}. In order to tame the chiral behavior of the
amplitudes containing the mass gap between the ground state and
excited heavy meson states $\Delta_{SH}$ we use the $1/\Delta_{SH}$
expansion of the chiral loop integrals~\cite{Fajfer:2006hi}.
As argued in Ref.~\cite{Becirevic:2006me} the $1/\Delta_{SH}$
expansion works well in an $SU(2)$ theory where kaons and etas,
whose masses would compete with the $\Delta_{SH}$ splitting, do not
propagate in the loops. Therefore we write down explicit expressions
for the chiral loop corrected Isgur-Wise functions specifically for
the strangeless states ($a=u,d$) in the $SU(2)$ theory:
\begin{eqnarray}
\xi_{aa}(w) &=&  \xi(w) \Bigg\{ 1 + \frac{3}{32\pi^2 f^2} m^2_{\pi} \log \frac{m^2_{\pi}}{\mu^2}  \Bigg[ g^2 2 (r(w)-1) \nonumber\\
&&\hskip -1cm - h^2  \frac{m^2_{\pi}}{4\Delta_{SH}^2} \left(1-w\frac{\tilde \xi(w)}{\xi(w)}\right) - h g \frac{m^2_{\pi}}{\Delta_{SH}^2} w(w-1)\frac{\tau_{1/2}(w)}{\xi(w)}\Bigg] \Bigg\},\nonumber\\
\label{eq:5}
\end{eqnarray}
and
\begin{eqnarray}
\tau_{1/2 aa}(w) &=&  \tau_{1/2}(w) \Bigg\{ 1 + \frac{3}{32\pi^2 f^2} m^2_{\pi} \log \frac{m^2_{\pi}}{\mu^2}  \Bigg[ - g\tilde g(2r(w)-1) - \frac{3}{2} (g^2+\tilde g^2) \nonumber\\
&&\hskip -2cm + h^2  \frac{m^2_{\pi}}{4\Delta_{SH}^2} \left(w-1\right) - h g \frac{m^2_{\pi}}{2\Delta_{SH}^2} \frac{\xi(w)}{\tau_{1/2}(w)} w(1+w) + h\tilde g \frac{m^2_{\pi}}{2\Delta_{SH}^2} \frac{\tilde \xi(w)}{\tau_{1/2}(w)} w(1+w) \Bigg] \Bigg\},\nonumber\\
\label{eq:6}
\end{eqnarray}
where
\begin{equation}
r(x) = \frac{\log(x+\sqrt{x^2-1})}{\sqrt{x^2-1}},
\end{equation}
so that $r(1)=1$ and $r'(1)=-1/3$. The first lines of Eqs.~(\ref{eq:5}) and~(\ref{eq:6}) contain the leading contributions while the calculated  $1/\Delta_{SH}$ corrections are contained in the second lines. Note that the positive parity heavy mesons contribute only at the $1/\Delta_{SH}^2$ order in this expansion since all the possible $1/\Delta_{SH}$ contributions vanish in dimensional regularization and the affected loop integral expressions have to be expanded up to the second order in $1/\Delta_{SH}$.

\par

\begin{figure}
\begin{center}
\hspace*{-0.4cm}\scalebox{0.8}{\includegraphics{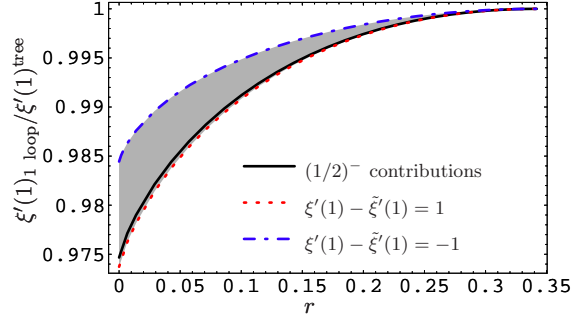}}
\end{center}
\caption{\label{plot_1} {\footnotesize\sl Chiral extrapolation of the slope of the IW
function at $w=1$ ($\xi'(1)$). Negative parity heavy states'
contributions (black line) and a range of possible positive parity
heavy states' contribution effects when the difference of slopes of
$\xi(1)$ and $\tilde \xi(1)$ is varied between $1$ (red dashed line)
and $-1$ (blue dash-dotted line).}}
\end{figure}

\begin{figure}
\begin{center}
\hspace*{-0.4cm}\scalebox{0.8}{\includegraphics{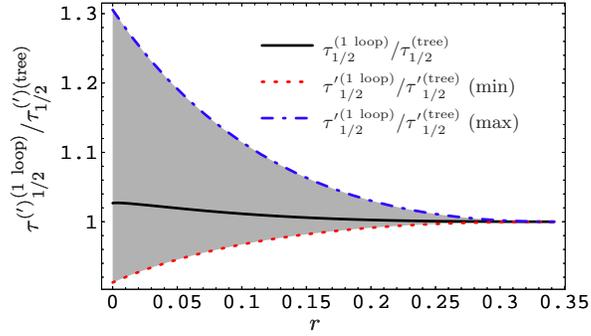}}
\end{center}
\caption{\label{plot_2} {\footnotesize\sl Chiral extrapolation of the $\tau_{1/2}$
function and its slope at $w=1$. $\tau_{1/2}(1)$ extrapolation
(black solid line) and a range of possible extrapolation effects of
$\tau_{1/2}'(1)$ (gray shaded region) when the difference of slopes
of $\xi(1)$, $\tilde \xi(1)$ and $\tau_{1/2}'(1)$ is varied between
$1$ (red dashed line) and $-1$ (blue dash-dotted line).}}
\end{figure}

We then plot the chiral behavior of the Isgur-Wise function
renormalization in the chiral limit below the $\Delta_{SH}$ scale in
FIGs.~\ref{plot_1} and~\ref{plot_2}.
\psfrag{xk1}[bc]{{$r$}}
\psfrag{xm1}[tc][tc][1][90]{{$\xi'(1)_{\mathrm{1~loop}}/\xi'(1)^{\mathrm{tree}}$}}
\psfrag{s1}[cl]{\footnotesize{$(1/2)^-$ contributions}}
\psfrag{s3}[cl]{\footnotesize{$\xi'(1)-\tilde \xi'(1)=1$}}
\psfrag{s4}[cl]{\footnotesize{$\xi'(1)-\tilde \xi'(1)=-1$}}
\psfrag{xk1}[bc]{{$r$}}
\psfrag{xm1}[cc][tc][1][90]{{${\tau^{(')}}^{\mathrm{(1~loop)}}_{1/2}/\tau^{(')\mathrm{(tree)}}_{1/2}$}}
\psfrag{s1}[cl]{\footnotesize{$\tau^{\mathrm{(1~loop)}}_{1/2}/\tau^{\mathrm{(tree)}}_{1/2}$}}
\psfrag{s3}[cl]{\footnotesize{${\tau'}^{\mathrm{(1~loop)}}_{1/2}/{\tau'}^{\mathrm{(tree)}}_{1/2}$
(min)}}
\psfrag{s4}[cl]{\footnotesize{${\tau'}^{\mathrm{(1~loop)}}_{1/2}/{\tau'}^{\mathrm{(tree)}}_{1/2}$
(max)}}
We have normalized the values of the extrapolated quantities at
$m_{\pi}\sim \Delta_{SH}$ to $1$ and perform the chiral
extrapolation using the Gell-Mann - Okubo formulae as in~\cite{Becirevic:2004uv}
$m^2_{\pi} = 8 \lambda_0 m_s r /f^2$ where $r=m_{u,d}/m_s$ and $8
\lambda_0 m_s / f^2 = 2m_K^2-m_{\pi}^2 = 0.468~\mathrm{GeV}^2$.
Presently no reliable estimates exist for the values of $\tilde
\xi'(1)$ and $\tau_{1/2}'(1)$, which feature in chiral extrapolation
involving opposite parity heavy states. Therefore we estimate their
possible effects by varying their relative values in respect to
$\xi'(1)$ between $1$ and $-1$ in our extrapolations. We see that
the effects of positive parity states' in the chiral loops on the
chiral extrapolation of $\xi'(1)$ appear to be mild (around one percent
in our estimate) below the $\Delta_{SH}$ scale (the gray shaded
region around the leading order result in black solid line).
Actually if $\xi'(1)-\tilde \xi'(1)$ is positive as reasoned in~\cite{Falk:1993iu}
and around $1$, these leading $1/\Delta_{SH}$ corrections almost vanish.
The same general chiral behavior can be attributed to $\tilde \xi'(1)$
with the substitutions $g \leftrightarrow \tilde g$, $\Delta_{SH}
\leftrightarrow -\Delta_{SH}$ and $\xi'(1) \leftrightarrow \tilde \xi'(1)$.
Also, the chiral extrapolation
(including small leading $1/\Delta_{SH}$ contributions) of the
$\tau_{1/2}(1)$ normalization appears fairly flat, indicating a
linear extrapolation as a good approximation, whereas the effects of
chiral loops on the extrapolation of its slope $\tau'_{1/2}(1)$
appear to be sizable, up to $30\%$ in our crude estimate.

\section{Discussion and Conclusion}

Within a HM$\chi$PT  framework, which includes even and odd parity
heavy meson interactions with light pseudoscalars as
pseudo-Goldstone bosons, we have calculated chiral loop corrections
to the functions $\xi$ and $\tau_{1/2}$. Motivated by the results of
Refs.~\cite{Becirevic:2006me,Fajfer:2006hi} where it was shown that
the leading pionic chiral logarithms are not changed by the
inclusion of even parity heavy meson states we consider chiral
extrapolation of Isgur-Wise functions. Our results are particularly
important for the lattice QCD extraction of the form factors. The
present errors on the $V_{cb}$ parameter in the exclusive channels
are of the order few percent. This calls for careful control over
theoretical uncertainties in its extraction~\cite{damir_fpcp07}. Our
results on the chiral corrections are
crucial in assuring validity of the form factor extraction and error
estimation coming from the lattice studies. Note also that from our
results given in Appendix one can deduce the chiral corrections in
the $B_s \to D_s$ decays which are not approached by experiment. Due
to the strange quark flavor of final and initial heavy meson states,
there is no leading pion logarithmic corrections making the lattice
extraction below the heavy meson parity splitting gap $\Delta_{SH}$
much simpler.

\par

In the $1/\Delta_{SH}$ expansion the opposite parity contributions
yield formally next-to-leading chiral log order corrections in a
theory with dynamical heavy meson fields of only single parity.
Therefore they compete with $1/\Lambda_{\chi}$ corrections due to
operators of higher chiral powers within chiral loops (yielding
contributions such as those of the central and right diagrams in
Figure~\ref{diagram_xi}), where $\Lambda_{\chi}$ is the
chiral symmetry breaking cut-off scale of the effective theory.
In a theory containing propagating heavy meson states of both
parities, the inclusion of such terms would in addition also yield
$1/(\Lambda_{\chi}\Delta_{SH})$ terms. Our present approach to the
estimation of the positive parity effects on the chiral
extrapolation is therefore valid with the assumption $\Delta_{SH} <
\Lambda_{\chi}$ where these additional contributions are further
suppressed.

\section*{Acknowledgment}
This work is supported in part by the European Commission RTN network, 
Contract No. MRTN-CT-2006-035482 (FLAVIAnet).
The work of S.F. and J.K. is supported in part by the Slovenian
Research Agency. J.O.E. is supported in part by the Research Council
of Norway. 

\appendix

\section{Complete expressions of 1-loop corrected Isgur-Wise functions}

Below are the complete expressions of the chiral 1-loop corrected
Isgur-Wise functions, calculated as explained in the text. For the
$\xi(w)$ we get
\begin{eqnarray}
\xi_{ab}(w) &=&  \xi(w) \Bigg\{ \delta_{ab} + \frac{1}{2} \delta Z_{2P_a(v')} + \frac{1}{2} \delta Z_{2P_b(v)}  + \frac{\lambda^i_{ac}\lambda^i_{cb}}{16\pi^2 f^2}   \nonumber\\
&& \times \Bigg[ g^2 \left((w+2) C_1(w,m,0,0) + (w^2-1) C_2(w,m,0,0)\right) \nonumber\\
&& - h^2 \frac{\tilde \xi (w)}{\xi(w)} \left( \sum_{i=1}^4 C_i (w,m,\Delta_{SH},\Delta_{SH}) + (w^2-w+1) C_2(w,m,\Delta_{SH},\Delta_{SH})  \right) \nonumber\\
&& - 2 h g \frac{\tau_{1/2} (w)}{\xi(w)} (w-1) \left(
C_1(w,m,\Delta_{SH},0) + w C_2(w,m,\Delta_{SH},0) + C_4(w,m,\Delta_{SH},0) \right) \Bigg]\Bigg\}.\nonumber\\
\end{eqnarray}
where the same formulae can be applied to $\tilde \xi (w)$ with the
substitution $g\leftrightarrow \tilde g$ and $\Delta_{SH}
\leftrightarrow -\Delta_{SH}$. For the $\tau_{1/2}(w)$ on the other
hand we obtain
\begin{eqnarray}
\tau_{1/2 ab}(w) &=&  \tau_{1/2} (w) \Bigg\{ \delta_{ab} + \frac{1}{2} \delta Z_{2P_a(v')} + \frac{1}{2} \delta Z_{2P_{0b}(v)}  + \frac{\lambda^i_{ac}\lambda^i_{cb}}{16\pi^2 f^2}  \nonumber\\
&& \hskip-1.5cm \times \Bigg[ g \tilde g \left((w-2) C_1(w,m,0,0) + (w^2-1) C_2(w,m,0,0)\right) \nonumber\\
&& \hskip-1.5cm - h^2 \left(w \sum_{i=1}^4 C_i (w,m,\Delta_{SH},-\Delta_{SH}) + (w^2-w+1) C_2(w,m,\Delta_{SH},-\Delta_{SH})  \right) \nonumber\\
&& \hskip-1.5cm + h g (w+1) \frac{\xi(w)}{\tau_{1/2} (w)} \left( C_1(w,m,0,-\Delta_{SH}) + w C_2(w,m,0,-\Delta_{SH}) + C_3(w,m,0,-\Delta_{SH}) \right) \nonumber\\
&& \hskip-1.5cm - h \tilde g (w+1) \frac{\tilde\xi(w)}{\tau_{1/2}
(w)} \left( C_1(w,m,\Delta_{SH},0) + w C_2(w,m,\Delta_{SH},0) +
C_4(w,m,\Delta_{SH},0) \right) \Bigg]\Bigg\}.\nonumber\\
\end{eqnarray}
In the above expressions $\delta Z_{2P} = (Z_{2P}-1)$ are the chiral
loop corrections to the heavy meson wavefunction renormalization:
\begin{eqnarray}
Z_{2P_a(v)} &=& 1 - \frac{\lambda^i_{ab}\lambda^i_{ba}}{16\pi^2 f^2}  \left [ 3 g^2 C'_1\left(0,m_i\right) - h^2 C'\left(\frac{\Delta_{SH}}{m_i} ,m_i\right)\right],\nonumber\\
\label{eq_Z_2_P}
\label{eq_Z2P}
\end{eqnarray}
for the negative parity doublet and
\begin{eqnarray}
Z_{2P_{0a}(v)} &=& 1 - \frac{\lambda^i_{ab}\lambda^i_{ba}}{16\pi^2 f^2}  \left[ 3 \tilde g^2 C'_1\left(0,m_i\right) - h^2 C'\left(-\frac{\Delta_{SH}}{m_i},m_i\right) \right],\nonumber\\
\label{eq_Z2S}
\end{eqnarray}
for the positive parity states. As in Ref.~\cite{Stewart:1998ke}, a
trace is assumed over the inner repeated index(es) (here $b$).

We make use of the $C_i$ loop integral  functions, of which
$C_{i}(x,m)$ have been defined in Ref.~\cite{Fajfer:2006hi}, while
$C_i(w,m,\Delta_1,\Delta_2)$ have been defined
in~\cite{Boyd:1995pq}. The $1/\Delta$ expansion of $C_{i}(x,m)$ has
been demonstrated in Ref.~\cite{Fajfer:2006hi} while for
$C_i(w,m,\Delta_1,\Delta_2)$ it follows as
\begin{align}
C_1(w,m,\Delta,0) =C_1(w,m,0,\Delta) &\to -(1/\Delta) C_1(m,0) - (1/\Delta^2) C_0(m) w + \mathcal O(1/\Delta^3), \nonumber\\
C_2(w,m,\Delta,0) =C_2(w,m,0,\Delta) &\to - (1/\Delta^2) C_0(m) + \mathcal O(1/\Delta^3), \nonumber\\
C_3(w,m,\Delta,0) =C_4(w,m,0,\Delta) &\to - (1/\Delta) C_1(m,0) + (2/\Delta^2) C_0(m) w + \mathcal O(1/\Delta^3), \nonumber\\
C_4(w,m,\Delta,0),C_3(w,m,0,\Delta) &\to \mathcal O(1/\Delta^3), \nonumber\\
C_1(w,m,\Delta,\Delta) = - C_1(w,m,\Delta,-\Delta) &\to (1/\Delta^2) C_0(m) + \mathcal O(1/\Delta^3), \nonumber\\
C_2(w,m,\Delta,\Delta),C_2(w,m,\Delta,-\Delta) &\to \mathcal O(1/\Delta^3), \nonumber\\
C_3(w,m,\Delta,\Delta),C_3(w,m,\Delta,-\Delta) &\to \mathcal O(1/\Delta^3), \nonumber\\
C_4(w,m,\Delta,\Delta),C_4(w,m,\Delta,-\Delta) &\to \mathcal
O(1/\Delta^3).
\end{align}
where
\begin{equation}
C_0(m)= -\frac{1}{4} m^4 \log \left(\frac{m^2}{\mu^2}\right).
\end{equation}
\bibliography{article}

\end{document}